\documentclass[12pt, onecolumn]{mn2e}
\usepackage[dvips]{graphicx}
\usepackage[dvips]{rotating}

\def\vcir{v_{\rm cir}}
\def\kms{{\rm \,km\,s^{-1}}}

\title[Satellites in the field and lens galaxies: SDSS/COSMOS vs. SLACS/CLASS]
{Satellites in the field and lens galaxies: SDSS/COSMOS vs. SLACS/CLASS}
\author[Jackson et al.]
{N. Jackson$^{1}$, S.E. Bryan$^{1}$, S. Mao$^{1}$, Cheng Li$^{2, 3}$\\
$^{1}$ Jodrell Bank Centre for Astrophysics, University of Manchester,
Alan Turing Building, Oxford Road, Manchester M13 9PL, UK\\
$^{2}$ Max-Planck-Institute for Astrophysics,
Karl-Schwarzschild-Str. 1, D-85741 Garching, Germany \\
$^{3}$ MPA/SHAO Joint Center for Astrophysical Cosmology, Shanghai
Astronomical Observatory, Nandan Road 80, Shanghai 200030, China\\
}
\begin{document}
\maketitle
\begin{abstract}
The incidence of sub-galactic level substructures is an important
quantity, as it is a generic prediction of high-resolution Cold 
Dark Matter (CDM) models which is susceptible to observational test.
Confrontation of theory with observations is currently in an uncertain
state. In particular, gravitational lens systems appear to show evidence
for flux ratio anomalies, which are expected from CDM substructures
although not necessarily in the same range of radius as observed.
However, the current small samples of lenses suggest that the lens 
galaxies in these systems are unusually often accompanied by luminous
galaxies. Here we investigate a large sample of unlensed elliptical
galaxies from the COSMOS survey, and determine the fraction of objects
with satellites, in excess of background counts, as a function of
satellite brightness and separation from the primary object. We find
that the incidence of luminous satellites
within 20~kpc is typically a few tenths of one percent for satellites of
a few tenths of the primary flux, comparable to what is observed for the
wider but shallower SDSS survey. Although the environments of 
lenses in the SLACS survey are compatible with this observation, the
CLASS radio survey lenses are significantly in excess of this.
\end{abstract}

\begin{keywords}
gravitational lensing - cosmology:galaxy formation
\end{keywords}

\Large

\baselineskip 12pt

\section{Introduction}

The current galaxy formation model is embedded in a cosmology
dominated by dark matter and dark energy (commonly referred to as the $\Lambda$CDM cosmology).
In this model, quantum fluctuations in the very early Universe are amplified by
gravitational instability dominated by cold dark matter. Further structure formation then follows
an hierarchical manner in which large structures are formed by the
agglomeration of smaller structures. Baryons can further condense
within nonlinear dark matter haloes to form visible galaxies.
The model has had a high degree of success in the description of
large-scale structures in the Universe (e.g. Percival et al. 2001; 
Tegmark et al. 2004; Cole et al. 2005; Eisenstein et al. 2005; 
Spergel et al. 2007).
.

Recently, the resolution of numerical simulations has allowed us to
address the question of the predicted structure on smaller,
galaxy-length, scales (Gao et al. 2004; Diemand et al. 2007). 
These simulations are difficult for two reasons;
firstly, the length scales are small enough that high resolution, and
consequently large numbers of particles, are required. Secondly, in the
central few kiloparsecs of galaxies, baryonic matter begins to dominate,
because it is able to lose energy non-gravitationally and cool into the 
centre of the gravitational potential wells produced by the dark matter.
This complicates simulations because additional non-gravitational
physics must be introduced, and this is typically done using either
hydrodynamical simulations or semi-analytic models.
It is important to confront the theoretical predictions with
observational evidence, because doing so effectively tests our
understanding of how galaxies form. Two lines of evidence are now
available.

The first body of evidence is provided by studies of our own Galaxy.
High-resolution simulations of a Milky-Way type halo (Moore et al. 1999,
Klypin et al. 1999, Gao et al. 2004, Diemand et al. 2007) predict that
the Milky Way should be surrounded by hundreds of dark matter
satellites down to circular velocities $\vcir \sim 20\kms$.
Until recently this was thought not to be the case, and efforts 
concentrated on finding mechanisms by which the satellites might have 
their star formation suppressed (e.g. Efstathiou 1992; Thoul \& Weinberg
1996; Gnedin 2000; Bullock et al. 2000) and thus not be
observable. In the last few years evidence of faint
satellites, many from the Sloan Digital Sky Survey (SDSS, York et al.
2000) has surfaced (e.g. Willman et al. 2005, Zucker et al. 2006, Zucker
et al. 2006, Belokurov et al. 2006, Belokurov et al. 2007) 
and the numbers are within a factor of a few of the
predicted number. However, recent ultra-high resolution simulations
indicate the number of subhaloes can reach tens of thousands
down to $\vcir \sim 3\kms$ (e.g. Madau et al. 2008; Springel et al. 2008),
and thus it is far from clear whether the predictions and observations
are consistent with each other.

The second avenue of attack is to use gravitational lensing of distant
galaxies and quasars, typically by foreground galaxies at $0.1<z<0.6$.
The effect of gravitational lensing is to produce multiple images of the
background object, and importantly, in a way that depends only on the
mass distribution of the lens and not on whether the mass is luminous or
not. In general, we can use constraints, usually in the form of lensed
image positions and fluxes, to solve an inverse problem and recover the
mass distribution of the lensing galaxy.

In the first instance, the usual procedure is to fit a smooth model for
the lens galaxy mass distribution. In some cases such a model fails to
fit the image positions, and in many cases it fails to fit the fluxes
(Kochanek 1991), of well-constrained lens systems. This problem was 
highlighted by Mao \& Schneider (1998) who attributed the failure to perturbations in the
smooth potential due to dark matter substructure at the level of a few
percent. Since then, the problem has been studied both for individual cases
and statistically for ensembles of lenses (Dalal \& Kochanek 2002;
Metcalf 2002; Chiba 2002; Kochanek \& Dalal 2004; Biggs et al. 2004).
The most secure evidence is obtained from lens systems with a
radio-loud source, since the image fluxes are not subject to
microlensing effects and (differential) dust extinction in this case, and systems with four images which
provide relatively large numbers of constraints. Dalal \& Kochanek
(2002) and Kochanek \& Dalal (2004) studied a sample of seven radio-loud 
lenses, mostly those discovered in the complete radio survey CLASS
(Myers et al. 2003, Browne et al. 2003) and concluded that the image 
flux anomalies when compared to
smooth models required a contribution of between 0.6\% and 7\% (90\%
confidence limit) from substructures, roughly as predicted by CDM simulations.

In comparison, the CDM simulations predict that typically 5-10\% of the total
halo mass is in substructures (e.g. Klypin et al. 1999; Moore et al. 1999;
Ghigna 2000). So far, so good; but there are indications that gravitational lens flux
anomalies do not provide a watertight vindication of CDM expectations.
First, the substructures are in the wrong place; the lenses constrain
matter distributions typically in the 5--10~kpc range, whereas the CDM
substructures are predicted mostly to lie in the dark-matter dominated
region further out (e.g. Mao et al. 2004; Xu et al. 2009). Secondly, when the
radio flux anomaly lenses are inspected more closely, it is often found
that ``substructure'' is visible, {\em and} luminous. Explicitly, bright
satellite galaxies which help to explain flux anomalies have been found
in MG0414+0534 (Schechter \& Moore 1993), 2016+112 (Schneider et al.
1985; Lawrence, Neugebauer \& Matthews 1993; More et al. 2009) and
CLASS~2045+265 (McKean et al. 2007) -- that is, in roughly half of the
known sample of four-image, radio loud lenses studied by Dalal \&
Kochanek (2002). In addition, the lens system CLASS~1608+656 contains a
complex lensing galaxy that is probably two galaxies in the process of
interaction (Jackson et al. 1998). In a previous paper, Bryan, Mao \& Kay (2008) studied
the Millennium Simulation (Springel et al. 2005) 
together with semi-analytical models to predict
the baryon distribution, and found that only 10\% of haloes with masses
larger than $10^{12}M_{\odot}$ were predicted to have bright satellite
galaxies within 14~kpc, and only 5\% within 7~kpc\footnote{Here and
elsewhere in the paper, we assume $H_0$=70~kms$^{-1}$Mpc$^{-1}$,
together with a matter density $\Omega_{\rm m}=0.27$, and a cosmological 
constant $\Omega_{\Lambda}=0.73$.}. Shin \& Evans
(2008) also addressed this problem by simulation of satellites within
galaxy-mass haloes, and found that the required total mass in satellites
to cause flux ratio anomalies, together with the distribution of
satellite masses, considerably underpredict the observed incidence of
large, luminous substructures. Both investigations thus paint a
consistent picture, the solution of which may be that luminous
substructures are more concentrated and hence more effective at
causing flux anomalies, or that the observed luminous substructures are
chance projections (Shin \& Evans 2008).

In this paper we extend this work to ask the following two questions. 
Firstly, what is
the incidence of bright satellites around field elliptical galaxies
which are {\em not} lenses? Are lens galaxies somehow untypical of the general
population, a hypothesis for which there is currently no evidence or
obvious explanation? Or does the explanation of some of the flux
anomalies by bright substructure indicate that flux anomalies are not
telling us about CDM substructure, which mostly lies at larger radius?
In this case we might expect to see that non-lens galaxies should have
bright substructure at the same rate as the lens galaxies. For this
work, we use the COSMOS survey (Scoville et al. 2007) which has HST
imaging of comparable quality of objects at comparable redshifts to the
CLASS lenses (Section 2).

Secondly, we compare the CLASS sample's incidence of satellites with 
that in the largest lens sample to date, the SLACS sample (Bolton et 
al. 2006, 2008). This sample, of generally lower-redshift lenses, is 
obtained by spectroscopic selection of luminous red galaxies from the 
SDSS which display more than one redshift 
system in their spectra, and has currently more than 80 reasonably 
certain lens systems. In addition, we also check the predicted fraction 
of field galaxies with satellites in the SDSS data in similar redshifts 
as SLACS lenses to see whether the SLACS optical lens sample is compatible
with field galaxies (Section 3). In Section 4 we briefly discuss the
results from these comparisons.

\section{Extraction of satellite galaxies}

\subsection{Treatment of the COSMOS images}

For this investigation, it is important to use images at similar resolution
to those which were used to image the lens galaxies -- that is, the
50-mas resolution of the Hubble Space Telescope (HST), corresponding to
about 300~pc at a redshift of 0.5. Previous work on the SDSS database
(e.g. Chen et al. 2006) yields information on scales of $>$1\farcs4, the
SDSS median resolution, which corresponds to about 10~kpc; this scale is
larger than the typical separation of observed CLASS satellites from the
lensing galaxy. A large sample of galaxies is also
needed for reasonable statistics; such a large, high-resolution sample
has recently become available in the form of the COSMOS survey of about
2 square degrees with the Advanced Camera for Surveys (ACS) on the HST
(Scoville et al. 2007; Capak et al. 2007). We therefore use the COSMOS 
data for this investigation, noting that its image quality has already
been proven to be useful in identifying lens systems (Faure et al.
2008; Jackson 2008), and that its distribution of redshift and magnitude
is broadly similar to the lens galaxies in the CLASS sample (Fig. 1).

\begin{figure}
\includegraphics[angle=-90,width=8cm]{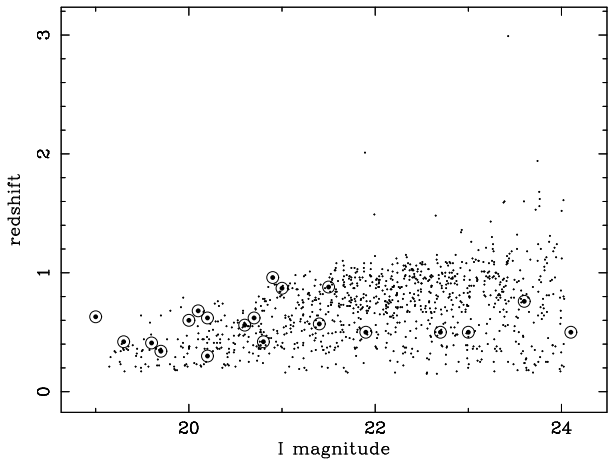}
\caption{Redshift vs. $I_{814}$ magnitude for COSMOS ellipticals (dots;
only every tenth object is plotted for clarity) and CLASS
lens galaxies (circled dots). In a few cases the CLASS redshift is
unknown and a value of 0.5 is assumed in this case.}
\label{magz}
\end{figure}

We use the ACS images in the F814W filter of the COSMOS field, which are
available from the COSMOS website (http://irsa.ipac.caltech.edu) as
reduced by Koekemoer et al. (2007). Early-type galaxies, similar to
the lens galaxies in gravitational lens systems, were selected
either by imposing a colour cut using the $B$ and $V$ photometry in
the COSMOS catalogue, or by using the {\sc tphot} parameter in the image
header to select those objects identified as elliptical galaxies by
requiring that 0$<${\sc tphot}$<$2. Only objects listed as good 
detections, not flagged as stars, with a photometric
redshift greater than 0.1 and with an $I_{814}$ magnitude greater than
24.9, corresponding to about 1 magnitude fainter than the LMC at $z$=0.5, 
were included\footnote{The stellar flagging was accomplished by
rejecting objects listed as stars, and also by rejecting all objects for
which the stellarity index was 1.00. Tests by eye on a selection of
objects not listed as stars showed that approximately three-quarters of
objects of stellarity index 1.00 were stars, and three-quarters of
objects of stellarity index 0.99 were galaxies. This is important
because it makes a significant difference (a few tens of percent) to the
background-subtracted counts, despite the relatively small numbers
involved (a few percent of the total sample).}
For each object, an image with radius of 11\farcs14
(corresponding to 20~kpc at $z>0.1$) was initially extracted.

For each image, the {\sc sextractor} program (Bertin \& Arnouts 1996)
was used to identify secondary objects in the aperture, and record any 
objects with $<$20~kpc distance from the primary and with a magnitude 
brighter than $I=24.9$.  Secondaries were rejected if they were considered to 
be cosmic rays, using a criterion based on the area of the object as 
fitted by {\sc sextractor}, and were also rejected if they were brighter
than the primary. Images of potential satellites were examined by eye
and rejected if they contained obvious artefacts, the usual reason being
proximity to the edges of the chips. The flux scale was based on the published
ACS/F814W magnitude zero-points and the calculations of Capak et al.
(2007), and we checked that the fluxes we recovered matched those in the
COSMOS catalogue where these overlapped, using a small adjustment from a
linear fit to measured and catalogue fluxes to improve agreement with
the COSMOS catalogue. Because the catalogue is based
on objects detected in multi-colour ground-based imaging, we could not
use it directly for detecting secondaries because close secondaries
would not be detectable given the poorer ground-based seeing. We could,
however, regard it as reasonable to assume that all primaries would be 
detected in the multi-colour catalogue.

\subsection{Background galaxy counts and the CLASS problem}


We must consider the possibility, for each detected secondary, that it
is a chance coincidence along the line of sight. To do this, 10000 random
positions within the COSMOS area were adopted. For each position, a
cutout was made in the same way as for the primary sample, and a 
centred, 3$^{\prime\prime}$-wide cutout of a random COSMOS object was added. 
This was done in order to reproduce the difficulty encountered in real
images of detecting faint satellites close to the primary COSMOS object.

All objects within 11\farcs1 (the angular distance corresponding to
20~kpc at a redshift of 0.1) were measured using the {\sc sextractor}
software and were visually inspected, in the case of objects close to
the central position, to check that they had not been introduced by the
insertion of the small cutout of the random COSMOS object. 
38715 secondaries were detected down to an I magnitude of
24.9. The results were then used to generate a grid of 
probability of detecting a random secondary, brighter than a certain 
magnitude and closer than a certain angular distance from a point
(Fig. 2). Fig. 2 also shows the secondaries found in the
CLASS lens systems (Table 1). 
Examination of archival HST images of these objects
using exposures of similar depths to COSMOS and extraction in the same
way with {\sc sextractor} yields the result that six secondaries 
(MG0414+0534=CLASS B0411+054, CLASS B0631+519, CLASS B1359+154,
CLASS B1608+656, CLASS B2045+265 and CLASS B2108+213) 
would have been detected had they 
lain within the COSMOS field, the remainder being too faint, too close
to the primary, or both. Although ACS images are not
available for some of the radio lens systems with satellites, existing
WFPC2 and other images (McKean et al. 2007; Schechter et al. 1993; 
Schneider et al. 1985) show that many of the satellites detected in
images of CLASS radio lens systems lie above the $I=24.9$ threshold, 
and therefore above the detection threshold of COSMOS. 

\begin{figure}
\includegraphics[width=8cm]{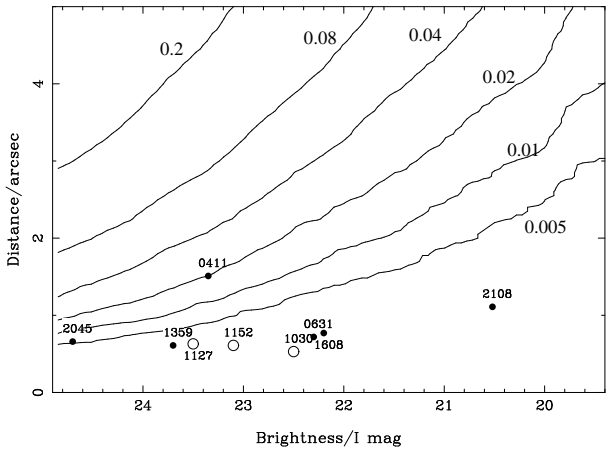}
\caption{Background counts in the COSMOS images. These are made using
10000 random pointings within the COSMOS catalogue, except avoiding the
edges of image frames. At each point in brightness-distance space, the
contour value represents the probability of obtaining a chance object
brighter and closer than the coordinates of that point. Satellites
detected in the CLASS survey (Table 1) are also plotted, labelled with
the right ascension of the CLASS object. Filled circles are the CLASS
satellites which are detected on HST images similar to those used to
produce the COSMOS images; open circles are those not detected.
}
\label{figbackground}
\end{figure}

In order to estimate the number of associated satellites in COSMOS
objects, we need to correct the counts around COSMOS images, to be discussed
below, for this possibility of chance coincidences along the line of sight.
Chen et al. (2006) and Chen (2009) comment on this problem of rejection
of ``interlopers''. Their analysis in the wider, shallower SDSS survey 
shows that using random positions can
potentially underestimate the number of interlopers, because due to
clustering of galaxies, random points are more biased towards voids 
than the sample of primary galaxies. In their simulations, a more
accurate estimate of the numbers of interlopers can be achieved by a
number of methods, including the estimation of the background using
points relatively near to primary galaxies. The distance required is
typically 400-500~kpc, corresponding to about 70\arcsec\ at a typical
COSMOS galaxy redshift. However, the median separation of the relatively
fainter COSMOS galaxies is about 50\arcsec, and it is hence likely that
the random-point estimation of interlopers gives a good estimate of the
background. (Incidentally, we note that underestimation of the background
would worsen the discrepancy between COSMOS and CLASS which is the major
result of this paper).

\subsection{COSMOS satellite counts and comparison with CLASS}


We first examine the statistics for the entire $z>0.1$ sample, using the
{\sc tphot} statistic to select the early-type galaxies as outlined
above. In Fig. 3 we show some examples of satellite objects 
found using this process.

\begin{figure*}
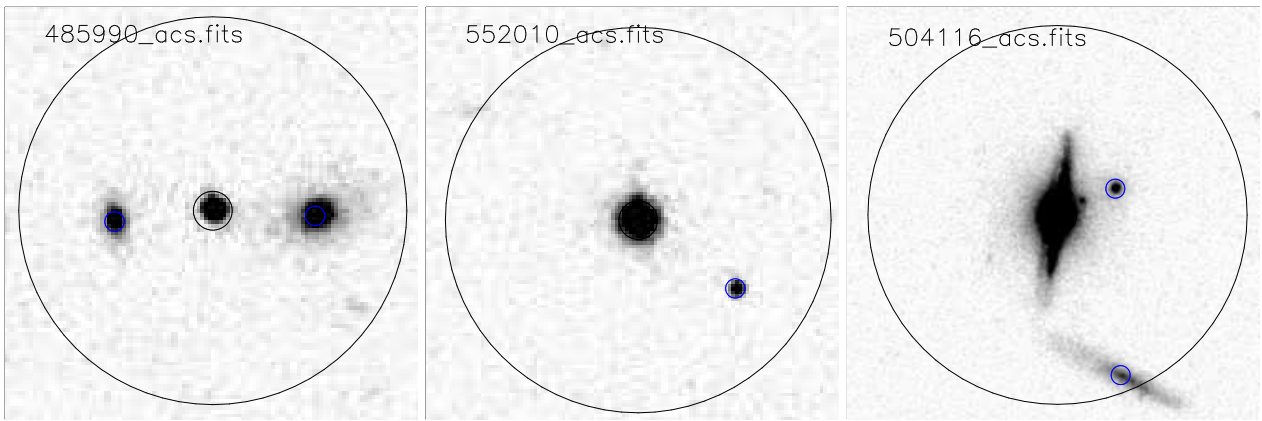

\centering
\includegraphics[angle=-90,width=5.5cm]{485990_acs.fits.ps}
\includegraphics[angle=-90,width=5.5cm]{552010_acs.fits.ps}
\includegraphics[angle=-90,width=5.5cm]{504116_acs.fits.ps}
\caption{Examples of satellites found within a 20-kpc projected radius
of primary galaxies in the COSMOS catalogue. In each case the large
circle shows the 20-kpc radius and circles indicate detected secondaries.}
\label{figexamples}
\end{figure*}

In the entire sample, 10974 early-type objects were examined and 46853 
satellites
identified before background subtraction, of which 4709 are within
20~kpc of the primary. The overall detection rate corresponds to a
signal of approximately 10$\sigma$ over background. 
Fig. 4 shows the results of this first
attempt. We describe the basic procedure, together with some of the
selection effects, before considering the results in more detail.

\begin{figure*}
\includegraphics[width=17cm]{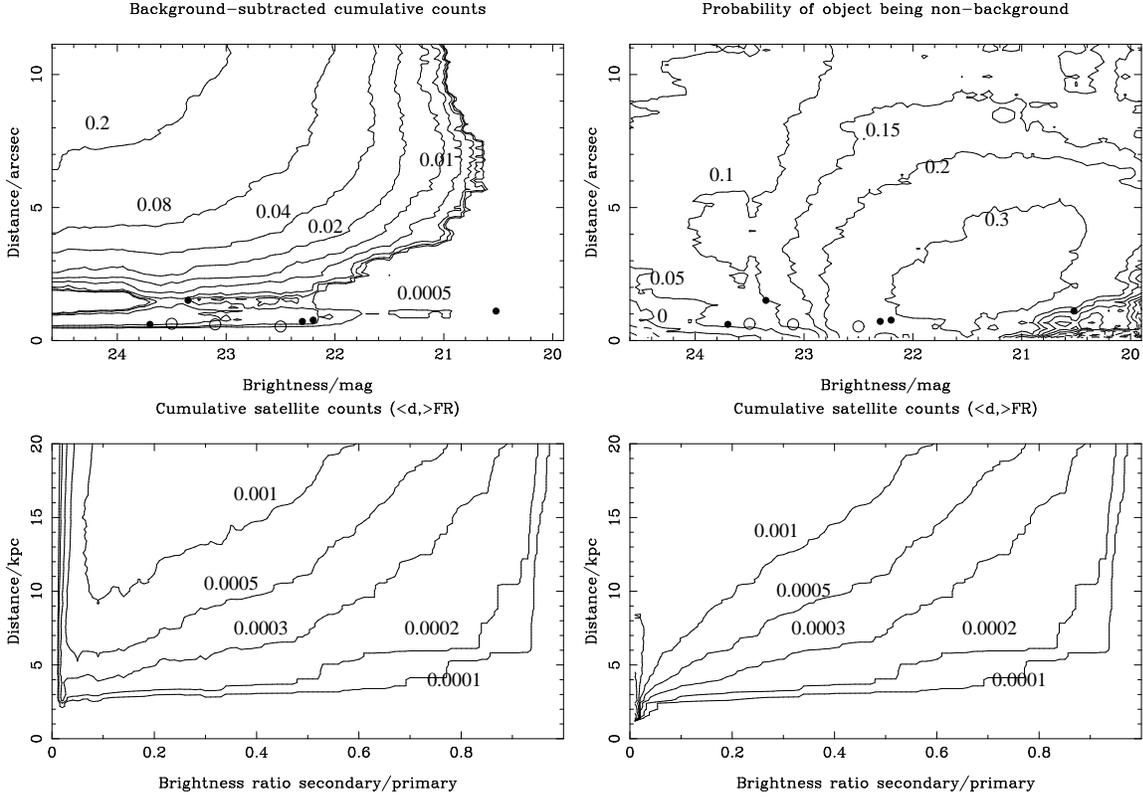}
\caption{Representations of the incidence of satellites in the COSMOS
survey objects, using objects identified as elliptical galaxies and with
photometric redshift $z_{phot}>0.1$. Top left: the contours at each grid
point represent the fraction of satellites brighter and closer than the
brightness and angular distance of that point, corrected for the
background counts. Top right: the probability, at each point in
brightness and primary-satellite distance, that a given satellite is
associated with the primary rather than being a background object. This
in general increases with increasing brightness and decreasing distance.
However, it becomes negative for weak satellites close to the primary,
because it is more difficult to detect such a satellite close to a
primary object than close to a random background point. Bottom:
cumulative satellite counts (as a fraction of objects with satellites
brighter and closer than a given flux ratio and linear distance). The
two plots are calculated by the ``minimalist'' and ``maximalist''
algorithms described in the text, and show that the counts become
unreliable at low flux ratios ($<0.1$) due to statistical uncertainties
associated with the detection limit. In three of the plots, filled dots
represent values for CLASS objects which have satellites found by the
SExtractor program, and circles represent CLASS satellites not found by
SExtractor.}
\label{figallcosmos}
\end{figure*}

We first calculated the background-subtracted cumulative
counts as a function of $I$-magnitude and angular distance. This was done
at each grid point ($f,d$) by calculating the fraction of primary objects which
have an observed satellite brighter than $f$ and closer than $d$,
together with the fraction of random background points which have
observed satellites brighter and closer than these limits, and
subtracting the two fractions. We also calculated the probability, at
each grid-point, of a detected satellite being associated with the
primary and not an interloper; this probability is smoothed in
Fig. 4 with a box size of 2\arcsec\ and 1 magnitude. 
The probability grid suffers severely from small-number statistics at low 
angular distances.

After background subtraction, we converted angular to linear distances
using the photometric redshift in the COSMOS catalogue, and magnitudes
to flux ratios between detected satellites and primaries. We also used
the smoothed fractional probability described above to weight the grid,
finally deriving cumulative satellite counts as a function of brightness
and separation (Fig. 4). This process is not unambiguous, because of the 
different ways that one can 
treat the detection limit. We approach this problem by adopting both a
``maximalist'' and ``minimalist'' algorithm, which will detect the
greatest and smallest possible number of satellites from the given data.
To illustrate this, suppose that we have six primary objects, three of which have a flux of
10 units and three of which have a flux of 5 units. A survey for
satellites is done with a detection threshold of 1 unit. In one of the
10-unit primaries, no satellite is detected, in the second a 1-unit
satellite is detected and in the third a 2-unit satellite is detected. A
similar arrangement is seen in the three 5-unit primaries.

The question is then: what is the fraction of objects which have
satellites brighter than 15\% of the primary flux? The ``minimalist''
approach would be to consider only the three primaries of flux 10 units,
in which a 15\% satellite could be seen, and to conclude that the answer
is 1/3. This, however, ignores the two 5-unit primary objects which
obviously have satellites brighter than 15\%; a ``maximalist'' approach
would then indicate a fraction of 3/5. A more strictly correct approach
would be to use any prior information about the distribution of galaxies
in a Bayesian analysis to find the most probable value. In practice,
what we do instead is to calculate fractions using the minimalist and
maximalist approaches. In any case where the results agree, we can be
sure of getting the correct answer. Not surprisingly, the results
diverge at small secondary:primary flux ratios where many secondaries
approach the $I=24.9$ detection threshold, and this typically limits our
analysis to flux ratios of about 2-2.5 magnitudes.

Both minimalist and maximalist approaches 
are plotted in Fig. 4, and areas of the grid in which the
two methods agree can be assigned an unambiguous value for the satellite
rate. Fig. 4 shows that the statistics are robust at flux ratios
of $\geq$0.1, or about 2.5 magnitudes. In Fig. 4 we also plot the
CLASS satellites. Having established that they are unlikely to be chance
background objects, this shows also that they are not representative of
the incidence of satellites in the COSMOS field galaxies. In particular,
three CLASS lens galaxy satellites with $f>0.1$ lie below the 0.1\%
probability contour on either algorithm.

We now estimate more formally the potential discrepancies between CLASS
and COSMOS satellites. There are two approaches we can take,
depending on whether or not we believe that the satellite in any
particular CLASS object is associated or not. If it is not associated,
we need to calculate the probability that a random pointing in the
COSMOS field has an interloper at the same (or greater) brightness and
the same (or closer) distance from
the pointing centre. If it is associated, we need to use the
background-subtracted COSMOS satellite probabilities, described in the
last section.

In CLASS B0631+519 we know that the satellite is not associated with the
primary, and in CLASS B1608+656 and CLASS B1359+154 we strongly suspect
on morphological grounds that the satellites are associated. For the
remainder of the CLASS objects with satellites, we do not know if the
satellite is associated. Table 2 shows the probability, for each CLASS
object with a satellite {\em that is detected by the setting of
SExtractor used for the COSMOS images}, of its occurring by chance, on
the assumption of non-association and on the assumption that it is
associated. Fig. 5 shows the six objects with satellites
detected by SExtractor, together with three objects from the COSMOS
database closely matched to them in flux density and photometric
redshift. There appears to be no statistically significant difference 
in the incidence of satellites between CLASS double and four-image
lenses (cf. Cohn \& Kochanek 2004) but the sample is currently very 
small.

We conservatively include CLASS B2114+022 as an object ``without'' satellites,
and take the more conservative probabilities of Table 2 where necessary.
We obtain the result that the six objects of Table 2, out of the 22 CLASS
objects, lie within a part of the distribution which has $<$2.1\% probability 
of occurring by chance. The binomial probability $P(\geq 6|p=0.021)$ is about
$2\times 10^{-5}$. If we instead use the fact that five satellites have a 
probability $<$0.007 of being background objects, or that they would have
been associated to a COSMOS object after background subtraction, we obtain
$P(\geq 5|p=0.007)=5.0\times 10^{-7}$.


\begin{figure*}
\includegraphics[width=15cm]{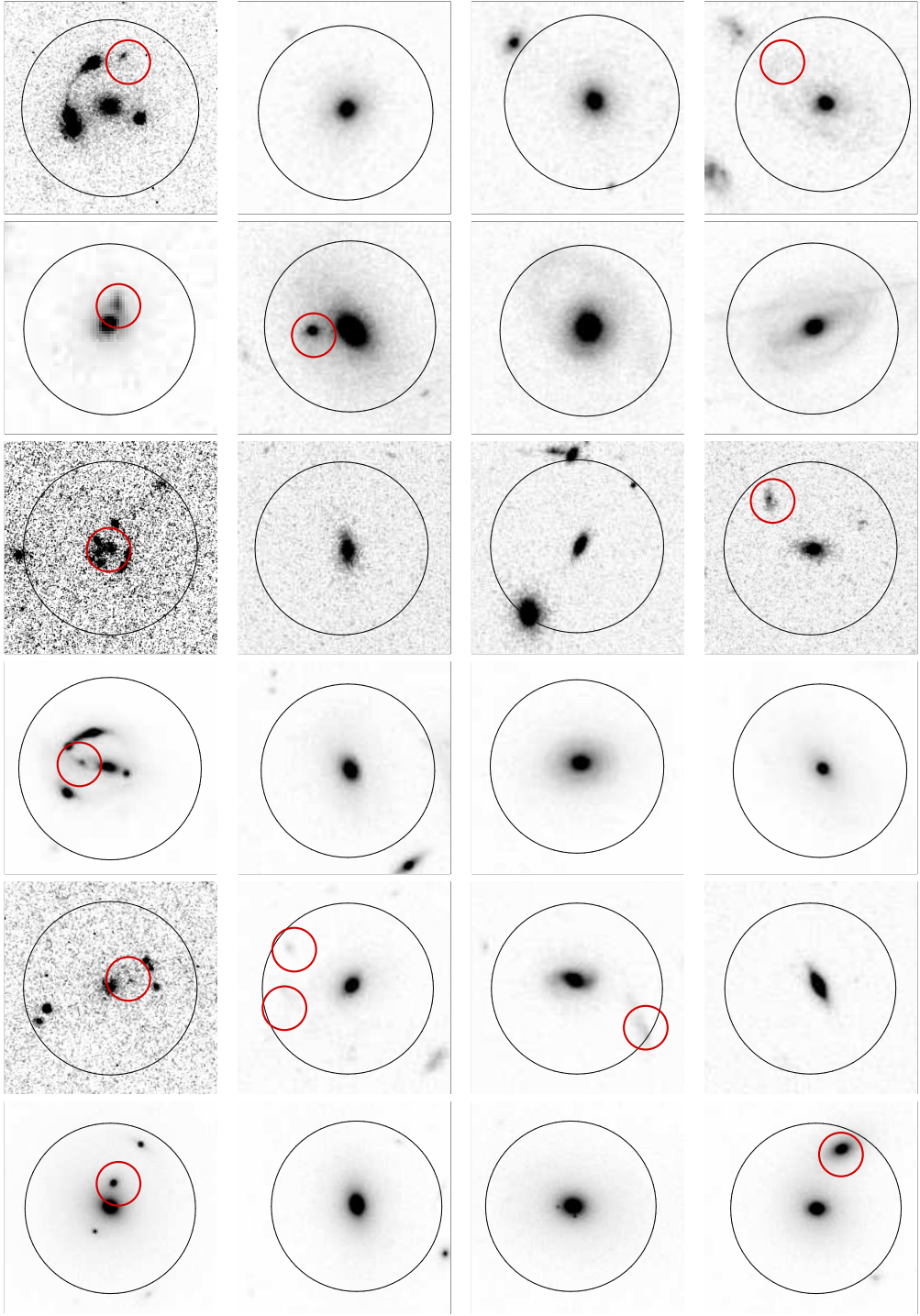}
\caption{Six CLASS lenses with detected substructure are shown in the
left-hand column (from top to bottom: MG0414+0534, CLASS B0631+519,
CLASS B1359+154, CLASS B1608+656, CLASS B2045+265, CLASS B2108+213). For
each lens, we have selected three objects from COSMOS matching it as closely as
possible in redshift and brightness. The figure illustrates satellites
that are found by circles; note that we have excluded lensed images from
the lens systems by this process. The large circle in each diagram has a
radius of 20~kpc. Only satellites fainter than the primary are
considered. Note that although satellites are found in COSMOS objects,
few are found close to the primary object in the way that seems to be
common in CLASS.}
\label{compfig}
\end{figure*}

\subsection{Limitations of the statistics}

There are a number of possible problems with the comparison of the CLASS
and COSMOS data:

\begin{enumerate}
\item The overall resolution limit of the survey, which is different in
linear units for objects of different redshift. For comparison with the
CLASS surveys, this does not matter because CLASS
data is taken using HST images\footnote{This is not strictly true, as
some of the CLASS images use the WFPC2 rather than the ACS.}, provided 
that we measure the HST images in the same way. 
\item The subtraction of the background is subject to small-sample
statistics at low distance from the primary.
\item The quasi-Bayesian problem of the detection limits, overcome as
described in the previous section.
\item The criterion 0$<${\sc tphot}$<$2 may not be the best way of
selecting early-type galaxies to compare with CLASS data.
\item The background statistics may affect the result.
\end{enumerate}

We show that in practice, these problems do not affect the result. This
is done by using only fainter objects, to assess effect ii); using a
colour cut ($B-V>0.7$) to assess effect iv); and using a different
realisation of the background to assess effect v). The results are
presented in Fig. 6 from which it is apparent that
the effect of these corrections is less than a factor of 2 in each case.

\begin{figure*}
\begin{tabular}{cc}
\includegraphics[width=8cm]{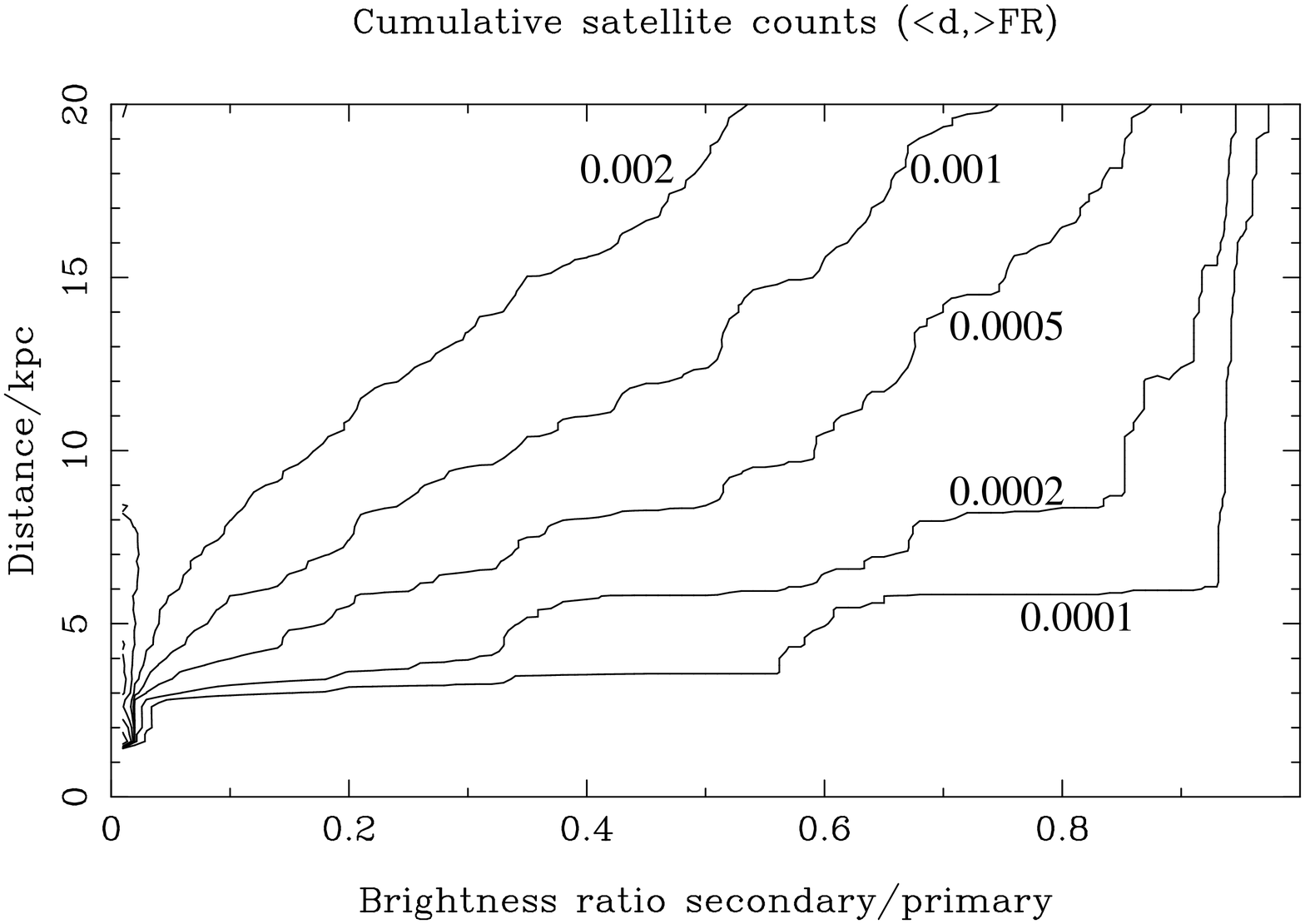}&
\includegraphics[width=8cm]{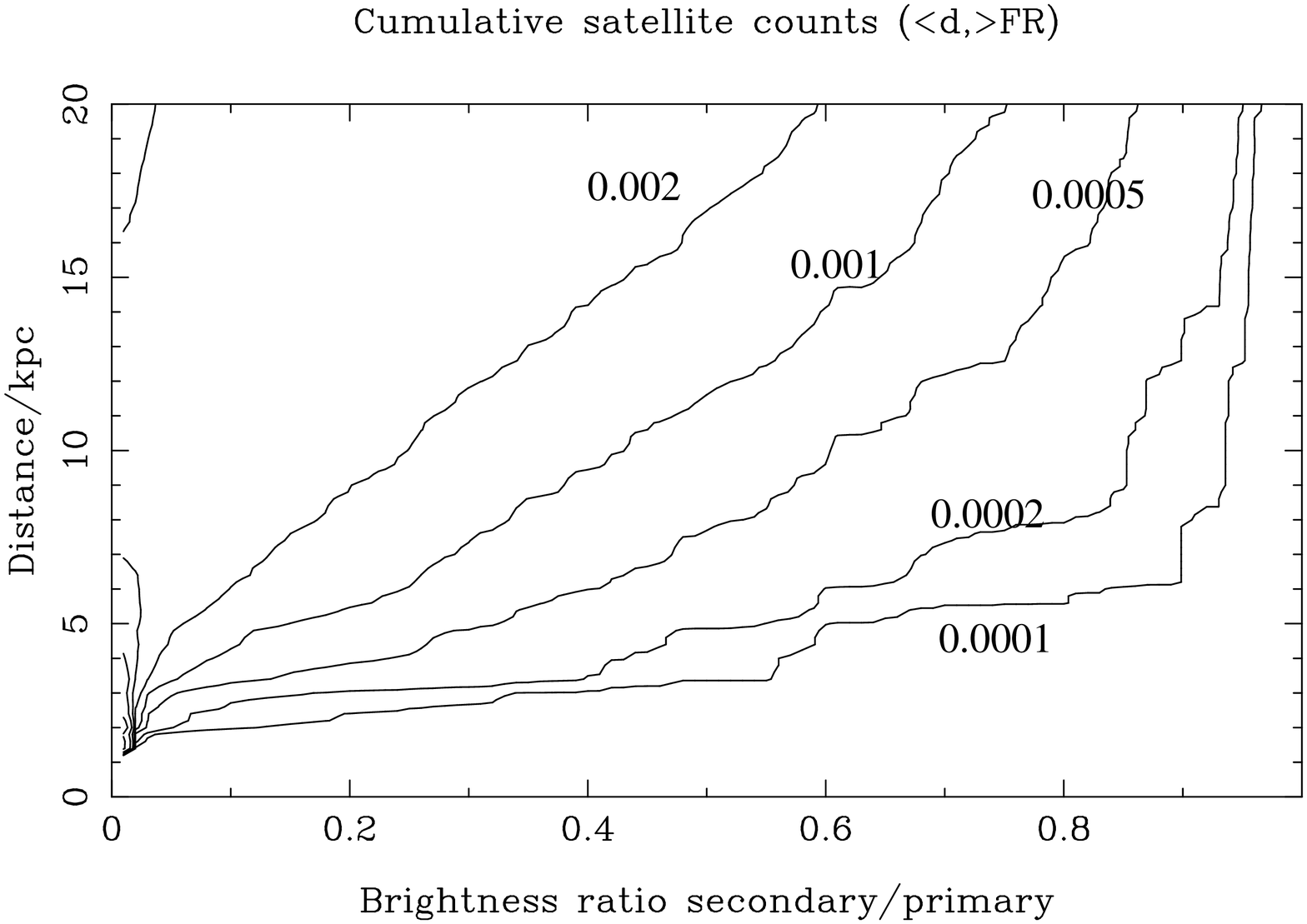}\\
\multicolumn{2}{c}{\includegraphics[width=8cm]{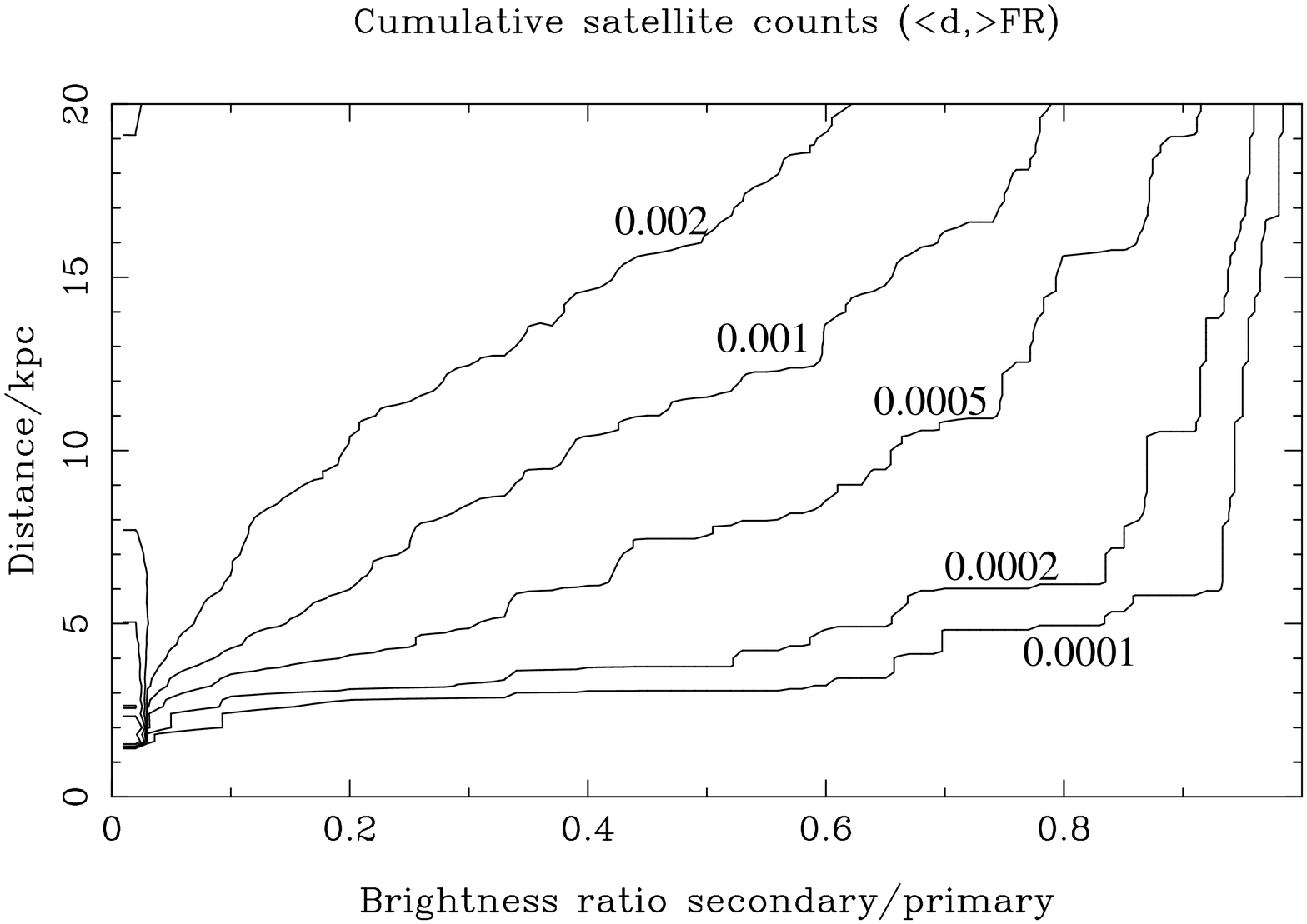}}\\
\end{tabular}
\caption{Cumulative background-subtracted satellite counts for the
COSMOS sample, under several different assumptions: (left) using a
different realisation of the background; (right) using a colour cut
instead of a galaxy classification; (bottom) using only fainter $I>20$
galaxies to assess the effect of removing the difficulty of finding
faint objects close to bright ones in the COSMOS data. The maximalist
counts are shown in each case.}
\label{figcompare}
\end{figure*}







\begin{table*}
\begin{tabular}{cccp{115mm}} \hline
Lens system & Primary/ & Primary/ & Comments/references \\
(\#images in& secondary & secondary &            \\
brackets)   &  flux ratio   & separation  & \\
            &           & (kpc)  & \\ \hline
CLASS B0128+437 (4) & - & - & No object visible within 20~kpc in the UKIRT
$K$-band image of Biggs et al. (2004).\\
CLASS B0218+357 (2) & - & - & No object visible within 20~kpc in HST image
(York et al. 2005).\\
MG0414+0534 (4) & 0.104 & 12.0 & Secondary identified by Schechter \& Moore
(1993) with $I=23.3$, close to one of the lensed images; this is 
detected by {\sc sextractor} on the HST/WFPC2 I-band image first
presented by Falco, Lehar \& Shapiro (1997).\\
CLASS B0445+123 (2) & 0.014 & 9.5 & Secondary is not identified by {\sc
sextractor} on the archival HST/ACS image (Proposal 9744, PI Kochanek)
due to faintness (and is in any case below the $I=24.9$ magnitude
limit).\\
CLASS B0631+519 (2) & 0.145 & 4.9 & Secondary on original discovery image
(York et al. 2005a) shown to be at a different redshift 
(McKean et al. 2004); however, as it is detected by {\sc sextractor} it
should be included in the statistics.\\
CLASS B0712+472 (4) & - & - & No secondaries within 20kpc in WFPC2/NICMOS images
(Jackson, Xanthopoulos \& Browne 2000)\\
CLASS B0739+366 (2) & - & - & No secondaries within 20kpc in archival WFPC2
image (PI Impey, proposal 8268)\\
CLASS B0850+054 (2) & - & - & No secondaries within 20kpc in archival
HST/ACS image (PI Kochanek, proposal 9744)\\
CLASS B1030+074 (2) & 0.100 & 3.5 & Secondary is visible 3.5~kpc from
primary galaxy in WFPC2 $I$-band image (Jackson, Xanthopoulos \& Browne
2000) but is not seen by {\sc sextractor} as a separate object.\\
CLASS B1127+385 (2) & 0.692 & 3.0 & HST/WFPC2 $I$-band image in discovery
paper (Koopmans et al. 1999) shows a secondary of $I$=23.5, about one
magnitude fainter than the primary. The secondary is detected by {\sc
sextractor} on the HST image, but with a very small area which would
have likely caused it to be rejected as a cosmic ray by the procedure
used to examine the COSMOS images.\\
CLASS B1152+199 (2) & 0.363 & 3.5 & This lens system consists of a bright
quasar lensed by an $I=19.6$ galaxy (Rusin et al. 2002) with a close
secondary, probably a satellite galaxy, 3.5 magnitudes fainter. {\sc
sextractor} does not detect the secondary, probably due to its
faintness and closeness to the primary.\\
CLASS B1359+154 (6) & 0.91 &  6.0 & The original discovery paper (Rusin et
al. 2001) shows the system to contain three lensing galaxies in a small
group. {\sc sextractor} detects all three in the HST image, and the
secondaries therefore appear in the statistics.\\
CLASS B1422+231 (4) & - & - & No secondaries are detected in high-quality
HST/ACS images (e.g. Impey et al. 1996) although in principle close
secondaries would stand a high chance of being hidden by bright quasar
images.\\
CLASS B1555+375 (4) & - & - & No secondaries are detected on archival WFPC2
$I$-band images (Proposal 8804, PI Falco) although the lensing galaxy
itself is very faint at $I\sim 24$.\\
CLASS B1600+434 (2) & - & - & The primary lens is an edge-on spiral galaxy
(Jaunsen \& Hjorth 1997) with a nearby, brighter galaxy which is,
however, $>$20kpc from the primary.\\
CLASS B1608+656 (4) & 0.191 & 4.9 & Deep HST/ACS images exist (Suyu et al. 2009)
which show clearly the two galaxies responsible for the lensing. {\sc
sextractor} fits these as distinct objects, and the fainter therefore
should be considered as a satellite. \\
CLASS B1933+503 (4) & - & - & No secondaries are detected on existing
HST/WFPC2 images. \\
CLASS B1938+666 (4) & - & - & No secondaries are detected on existing 
HST/WFPC2 images (King et al. 1998). \\
CLASS B2045+265 (4) & 0.05 & 5.1 & A secondary galaxy is detected by McKean
et al. (2007) on adaptive-optics Keck images. It is also seen on
archival HST/WFPC2 images and fitted by {\sc sextractor}.\\
CLASS B2108+213 (2) & 0.076 & 5.6 & A secondary galaxy is detected by McKean
et al. (2007) on HST/ACS images and also fitted by {\sc sextractor}.\\
CLASS B2114+022 (2?) & - & - & This system consists of two galaxies (Augusto
et al. 2001) at redshifts 0.32 and 0.59; it is not clear what the lens
configuration is here, and we do not include the object in the statistics.\\
CLASS B2319+051 (2) & - & - & No close secondaries are detected on existing
images; a secondary galaxy G2 was imaged by Rusin et al. (2001) but is
just outside the 20-kpc radius.\\ \hline
\end{tabular}
\caption{Details of secondary objects near the lensing galaxies of CLASS
objects, taken from the literature. $I$-band observations have been used
for the flux ratios; in a few cases, where the lens redshift is unknown,
a value of $z=0.5$ has been assumed.}
\end{table*}

\begin{table}
\begin{tabular}{lcc} \hline
Object & Prob. (Background) & Prob (Associated) \\ \hline
MG0414+0534 & 0.021 & 0.003 \\
CLASS B0631+519 & 0.002 & (0.002) \\
CLASS B1359+154 & (0.003) & 0.0001 \\
CLASS B1608+656 & (0.002) & 0.0002 \\
CLASS B2045+265 & 0.007 & 0.0006 \\
CLASS B2108+213 & 0.001 & 0.0005 \\ \hline
\end{tabular}
\caption{For each CLASS object with a satellite detected by SExtractor,
we show the probability of such a satellite being a random line-of-sight
superposition in the COSMOS field, and the probability of a COSMOS
object having a secondary brighter and closer than the CLASS object on a
background-subtracted analysis (see text). Brackets indicate excluded
cases; for example, CLASS B0631+519's secondary is known to be at a
discordant redshift and therefore unassociated.}
\end{table}

\begin{figure*}
\centering
\includegraphics[width=8cm]{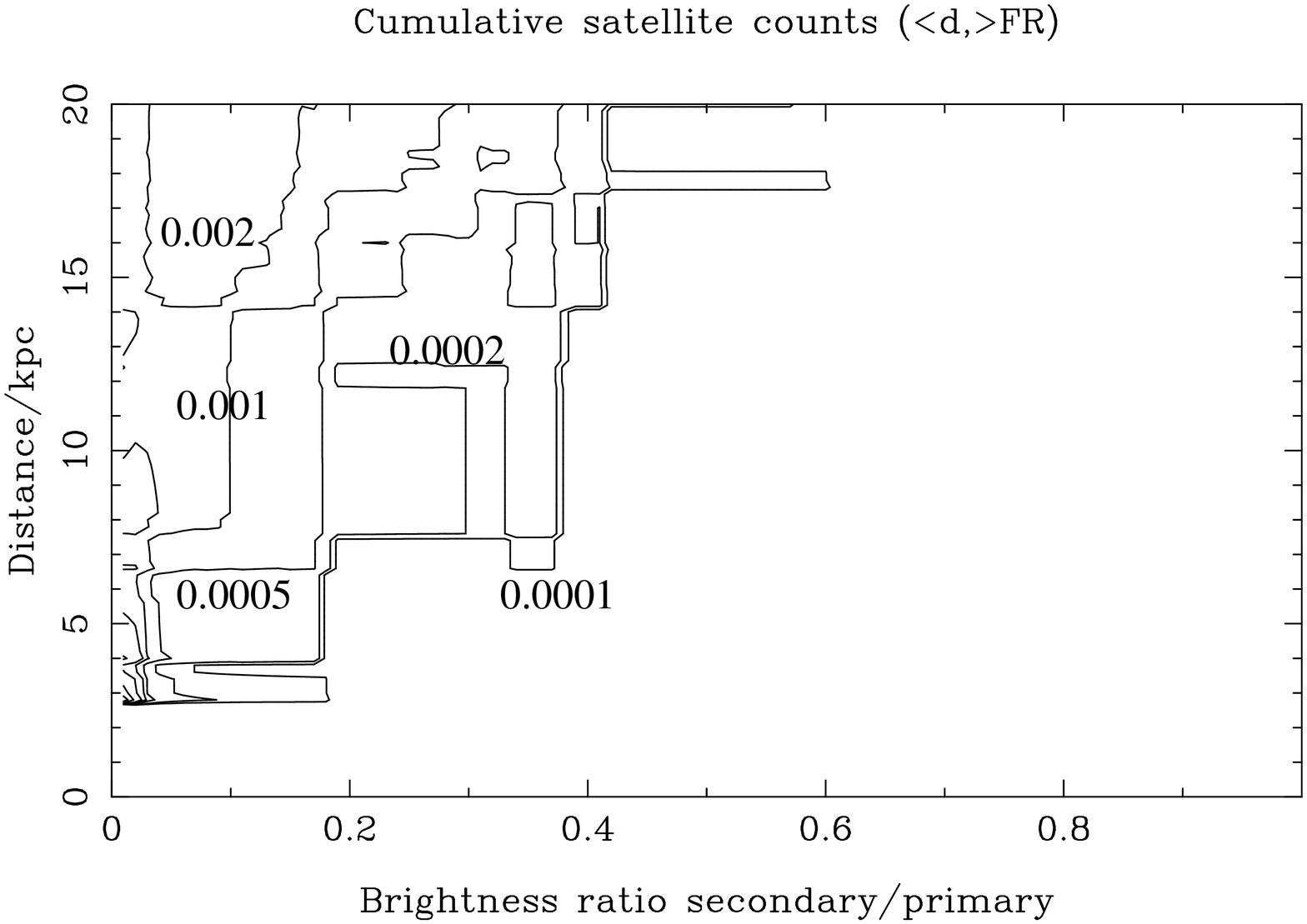}
\includegraphics[width=8cm]{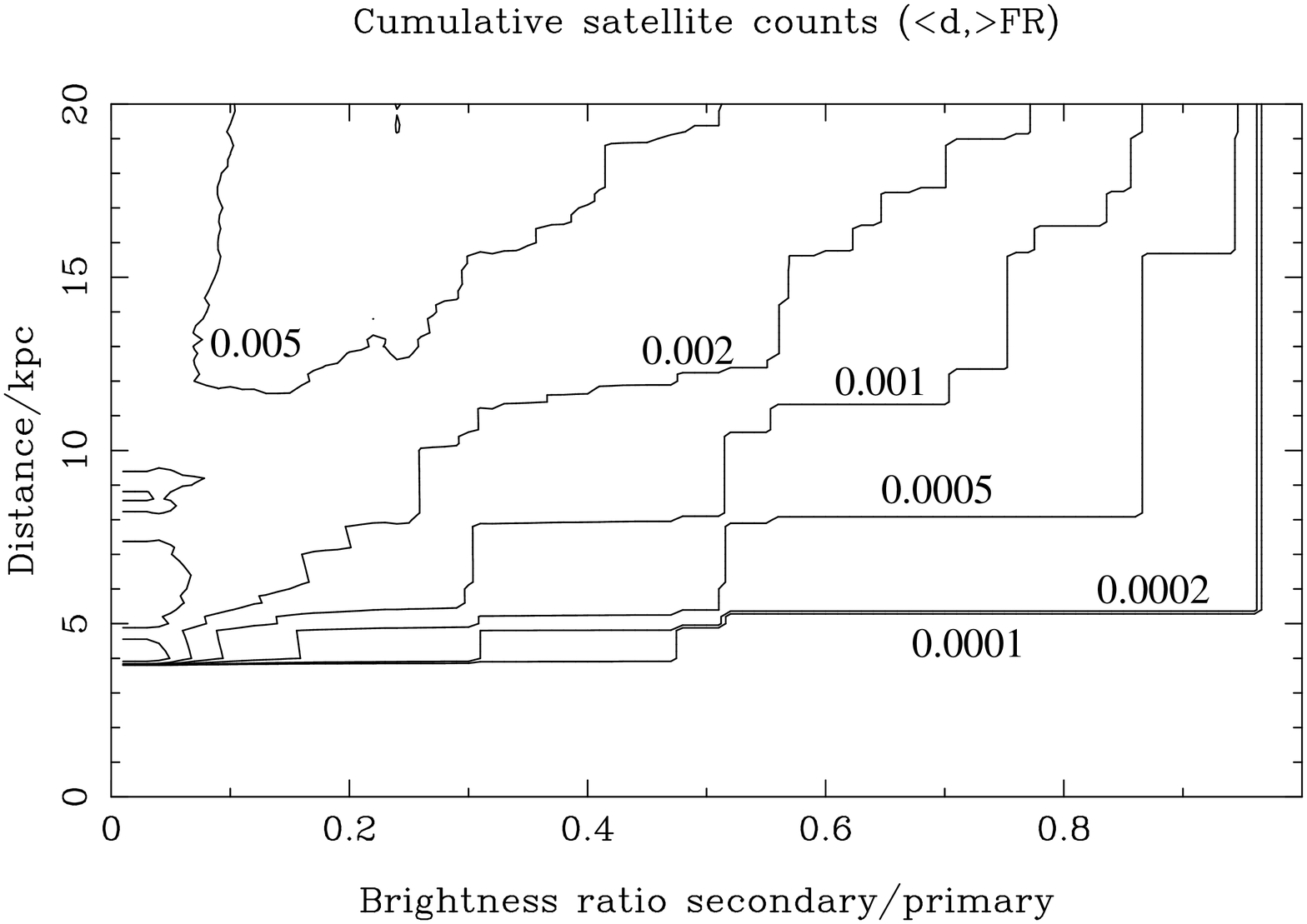}
\label{cosmossim}
\caption{Cumulative background-subtracted COSMOS counts for two redshift
ranges, 0.4$<z<$0.6 and 0.9$<z<$1.1}.
\end{figure*}

\begin{figure*}
\begin{tabular}{cc}
\includegraphics[angle=-90,width=8cm]{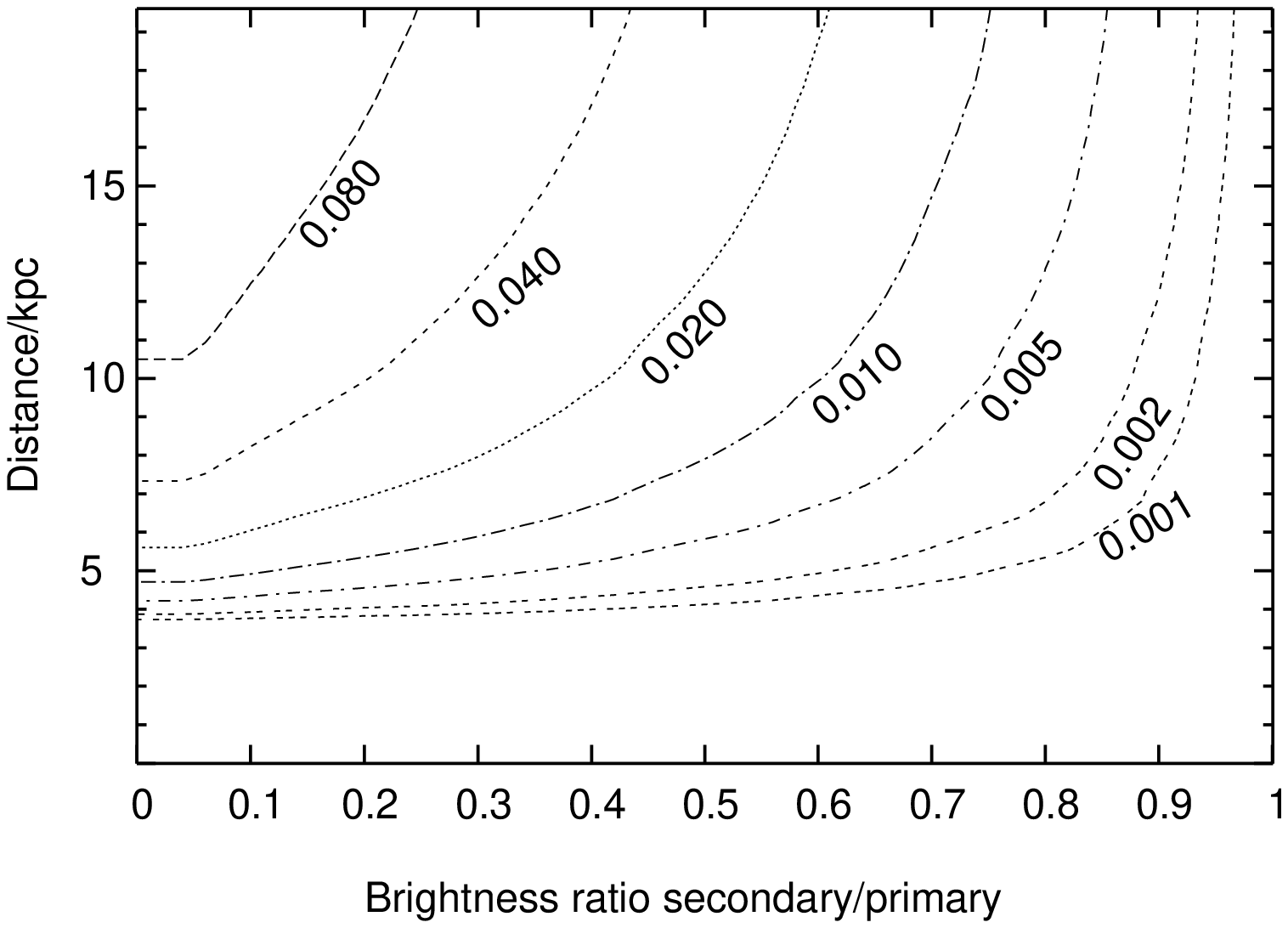}&
\includegraphics[angle=-90,width=8cm]{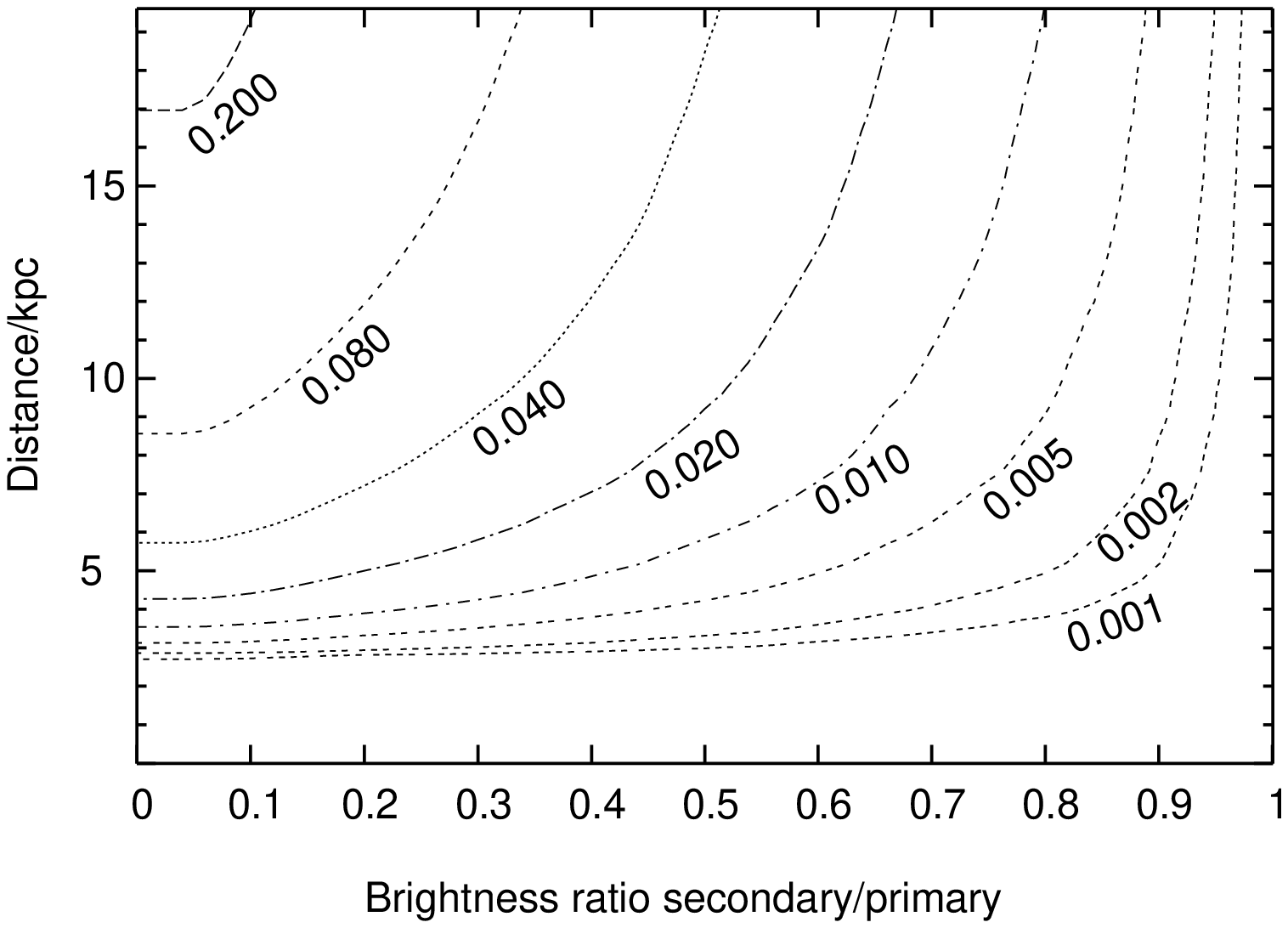}\\
\end{tabular}
\caption{(Top) The contours show the proportion of galaxies which contain
satellites at less than a certain distance $d$, and brighter than a
flux ratio $f$, from the primary, using the Millennium Simulation and the
method of Bryan et al. (2008). 
At each grid point, the contoured quantity represents the proportion of
primary galaxies which have secondaries at flux ratio $f$ or brighter,
at a distance $d$ or closer. The left-hand plot shows the results for all galaxies
with redshifts $0.4<z<0.6$, and the right-hand plot for $0.9<z<1.1$.}
\label{cosmosz}
\end{figure*}

It is also interesting to compare these results with simulations and to investigate whether
the fraction of satellite galaxies varies with redshift.  We can do this easily by
considering two samples of galaxies, 0.4 $<$ z $<$ 0.6 and 0.9 $<$ z $<$ 1.1, obtaining
two separate results for low- and high redshift galaxies which are shown in
Fig. 7.

For comparison, we have (as in Bryan et al. 2008) used the De Lucia and Blaizot
(2007) semi-analytic models run on the Millennium Simulation.  Haloes were selected
from the galaxy catalogue by imposing a minimum mass cut of $10^{12} h^{-1}$ M${_\odot}$. 
All galaxies satisfying the same cutoffs in brightness and colour selection (discussed
before) were considered.  We search within the virial radius of each halo for a
companion galaxy.  While Bryan et al. (2008) considered only central galaxies,
we have included all galaxies satisfying the imposed cuts.  The sample is however,
still dominated by central objects due to the small number of satellites satisfying
the mass cuts. 

It can be seen (Fig. 8) that, for the mass range considered, the simulations give typical
satellite fractions of about 8 times that of the COSMOS values, but even then
underpredict the satellite frequency observed in the CLASS samples. However, the 
satellite population consists of exclusively ``orphan'' galaxies which have been
stripped of their dark matter haloes (see Bryan et al. 2008 for further discussion). 
The fraction found in simulations does however depend on the lower mass limit imposed 
when selecting host haloes; higher mass haloes are more likely to host a companion
galaxy.  Increasing the mass cut to $10^{13} h^{-1}$ M${_\odot}$ increases the
fraction by a factor of $\sim$ 2.  We find we are able to reproduce the number
density of the COSMOS sample by imposing a minimum mass cut of 
$ \sim 3 \times 10^{11} h^{-1} $ M${_\odot}$ (on non-central galaxies) , in
doing so our fraction of galaxies found to host a companion is reduced by less
than a factor of two (i.e. the fractions are comparable within the level of
uncertainty of the observations).


\section{Neighbour counts around early-type field and lens galaxies at low redshifts}

In this section, we study the fraction of galaxies with satellites for
both field and lens galaxies selected from SDSS. Both sets of galaxies
are selected from redshift 0.05 to 0.4.

\subsection{Satellites in field galaxies}
\label{sec:sdss}

For  comparison, we have  performed a  similar analysis  on a sample of
galaxies in the low-redshift Universe drawn from the SDSS.  
For  this we take data from {\tt
  Sample  dr72} of  the New  York University  Value  Added Catalogue
(NYU-VAGC).   This  is  an  update  of the  catalogue  constructed  by
Blanton et al. (2005b),  is  based  on  the  seventh data  release  of  the
SDSS (DR7; Adelman-McCarthy et al. 2009), and  is publicly available on
the    NYU-VAGC   website\footnote{http://sdss.physics.nyu.edu/vagc/}.
Starting  from  {\tt  Sample  dr72},  we select all galaxies
with $r <18$ and spectroscopically measured redshifts $z<0.5$.  Here
$r$  is  the  $r$-band  Petrosian apparent  magnitude,  corrected  for
Galactic  extinction.  We then trim this sample so that it has roughly
the same distribution in both redshift and stellar mass as the SLACS
sample. This gives rise to a final sample of 75839 galaxies.
Detailed description of the NYU-VAGC can be found in Blanton et al. (2005b)
and
the methodology of estimating the stellar mass for SDSS galaxies is
described in Blanton \& Roweis (2007).

In  order to  count the  companions  around these  galaxies, we  first
construct a {\em photometric}  reference sample, from {\tt datasweep}
of the NYU-VAGC (version {\tt dr7}, see the NYU-VAGC website),
by selecting all galaxies with $r<21$. The resulting sample
includes $\sim$26 million galaxies.  We  then construct a  set of  10 random
samples that have  the same selection effects as  the reference sample
(a detailed account of the observational selection effects accompanies
the NYU-VAGC  release).  For each  object in our sample, we count
the number of companions in  the reference sample,
and  in each  of  the random  samples,  within a  given  value of  the
projected radius  $R_p$ and with  a minimum value of  luminosity ratio
$r_{L}^{min}$, determined  by the difference in  the $r$-band apparent
magnitude  between the  companions and  the galaxy  in  question.  The
average correlated  neighbour counts $N_c$ per galaxy,  as functions of
$R_p$ and $r_{L}^{min}$, are then  given by the difference between the
result from the  reference sample and the average  one from the random
samples.   By evaluating  and  subtracting the  counts  in the  random
samples, we  can make a statistical correction  for chance projections
of   foreground   and  background   galaxies   that   lie  along   the
line-of-sight.  This method is similar to that used in Li et al (2006),
Li et al. (2008a,b),   where  the   authors  computed
neighbour counts  around active  galaxy nuclei (AGN)  and star-forming
galaxies  (SFGs)  from  SDSS  and  examined the  connections  of  star
formation  and AGN activity  with tidal  interactions.  The  reader is
referred  to those  papers for  detailed description  and tests  of this
method.

The  result is  shown in  Fig. 9,  where we  plot the  contours
of  the
average correlated neighbour counts  around our early-type galaxies in
the plane of $R_p$ and $r_L^{min}$. It can  be seen
from the figure that the SDSS fractions are comparable to those
  seen in SLACS (and COSMOS), within the statistical noise.

\begin{figure}
\includegraphics[width=\columnwidth]{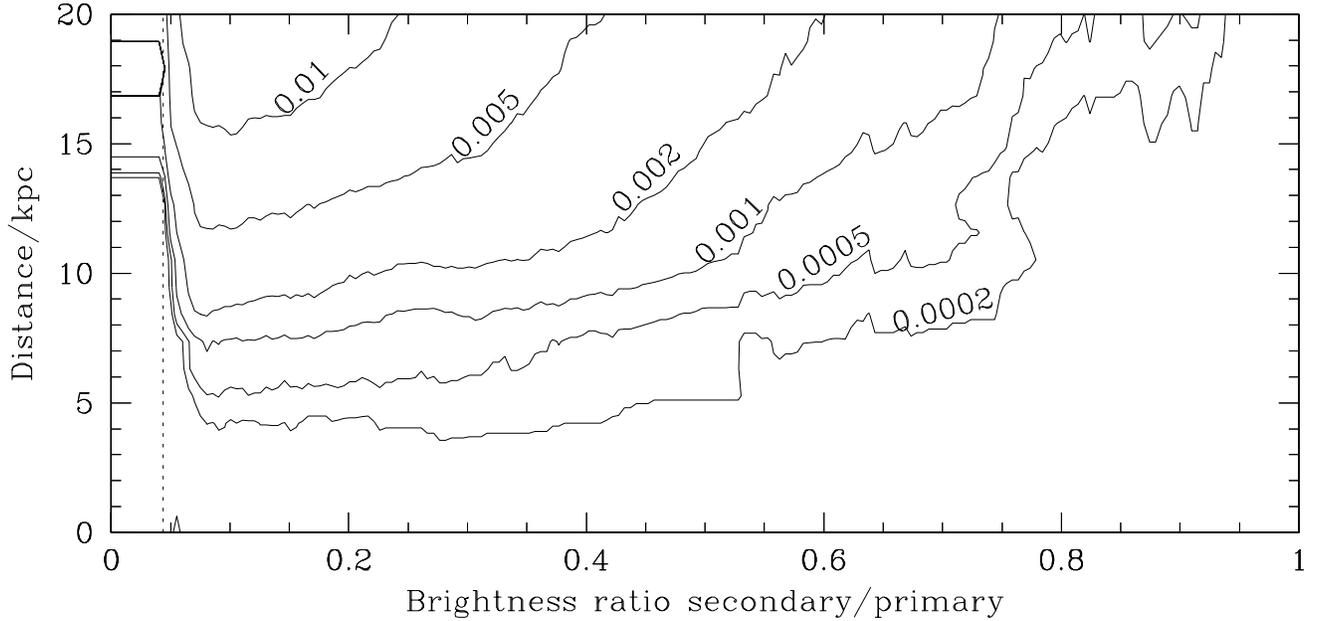}
\caption{Cumulative detection rate of SDSS satellites, as a function of
secondary/primary flux ratio and linear distance.}
\label{figsdss}
\end{figure}

\subsection{Satellites in SLACS lenses}

The major satellite anomaly detected to date has been the excess of
light substructure in CLASS lenses compared to normal elliptical
galaxies. However, the largest survey for gravitational lenses to date
has been the SLACS survey (Bolton et al. 2006) which targets luminous
red galaxies in the SDSS 
and examines the SDSS data for systems which contain more
than one spectroscopic redshift within a $\sim 3^{\prime\prime}$ fibre.
The intrinsic efficiency of this procedure has resulted in the discovery
of nearly 100 lens systems. All of these are systems in which a
background galaxy is imaged by a foreground galaxy; because of the SLACS
selection of bright objects, most of the systems have a lens redshift of
0.1--0.3, somewhat lower than that of CLASS.

Fields of the 64 most certain SLACS lenses from the most recent
compilation (Bolton et al. 2008; lens category YES) have been
examined; in most cases these fields extend to 20~kpc, or nearly
20~kpc, from the centre of the lens galaxy. SExtractor has again been
used to detect all satellites out to a 7\arcsec\ radius, 
with the exception of objects in the
regions covered by source structure; these regions are identified by
SLACS mask images and can be blanked out. They do not significantly
affect the statistics, as most such regions are within about 5~kpc of
the lens galaxy (Fig. 10). Fig. 11 shows the satellites
in the SLACS images as a function of linear and angular distance.

\begin{figure}
\includegraphics[angle=-90,width=8cm]{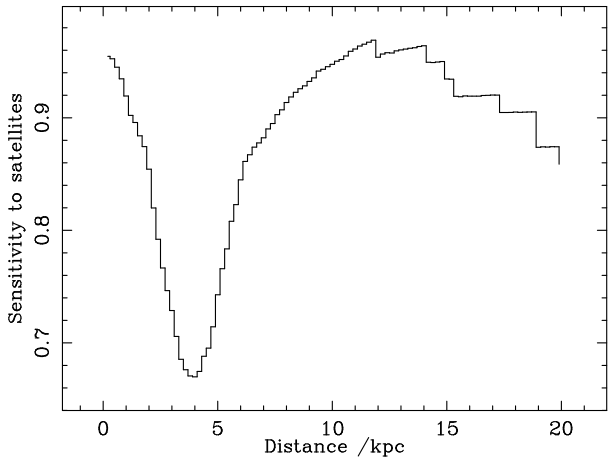}
\caption{Sensitivity to SLACS satellites as a function of linear
distance, due to masking of lensed source structure and the edges of the images}
\label{figslacssensitivity}
\end{figure}

\begin{figure*}
\centering
\includegraphics[width=8cm]{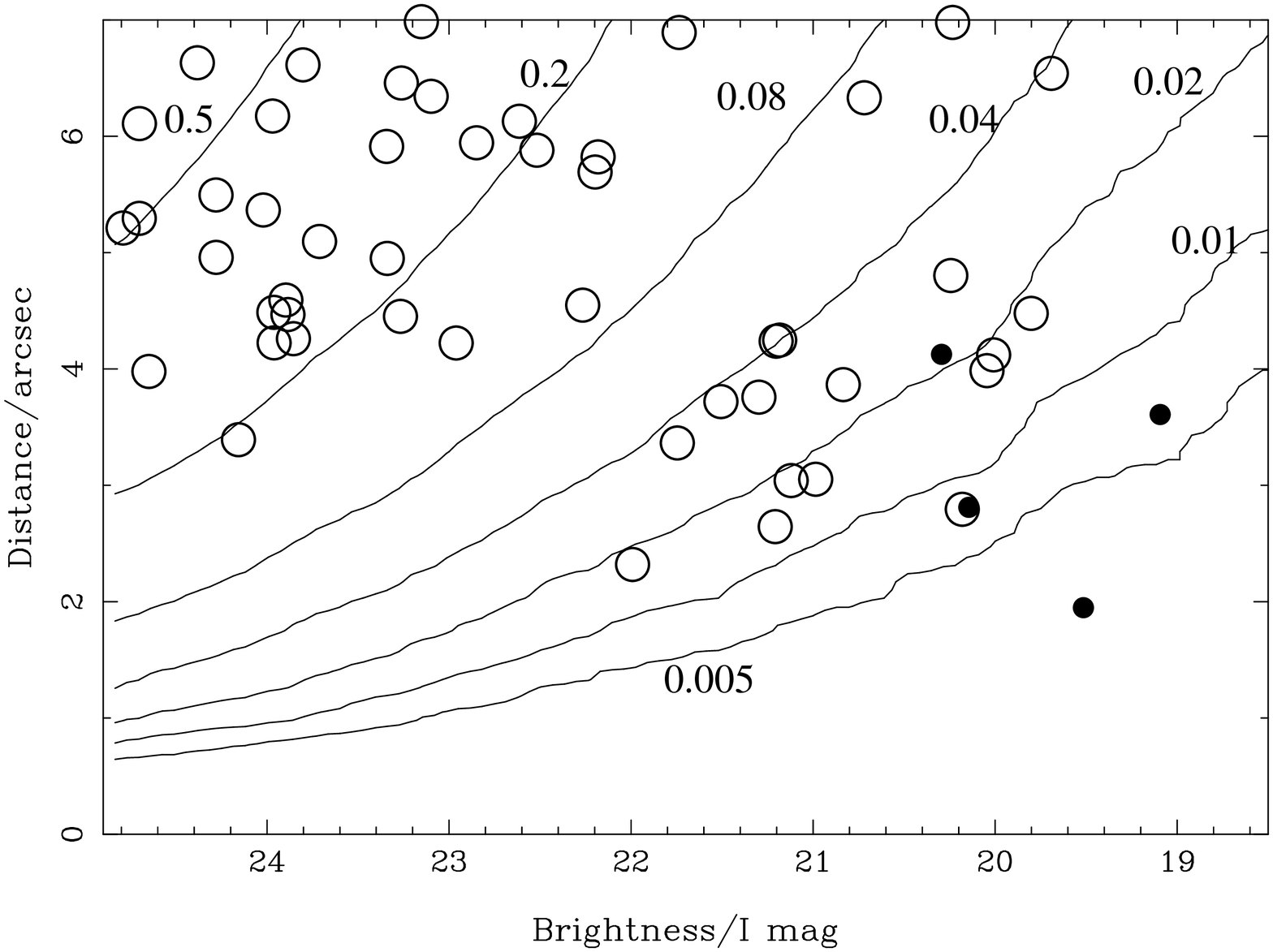}
\includegraphics[width=8cm]{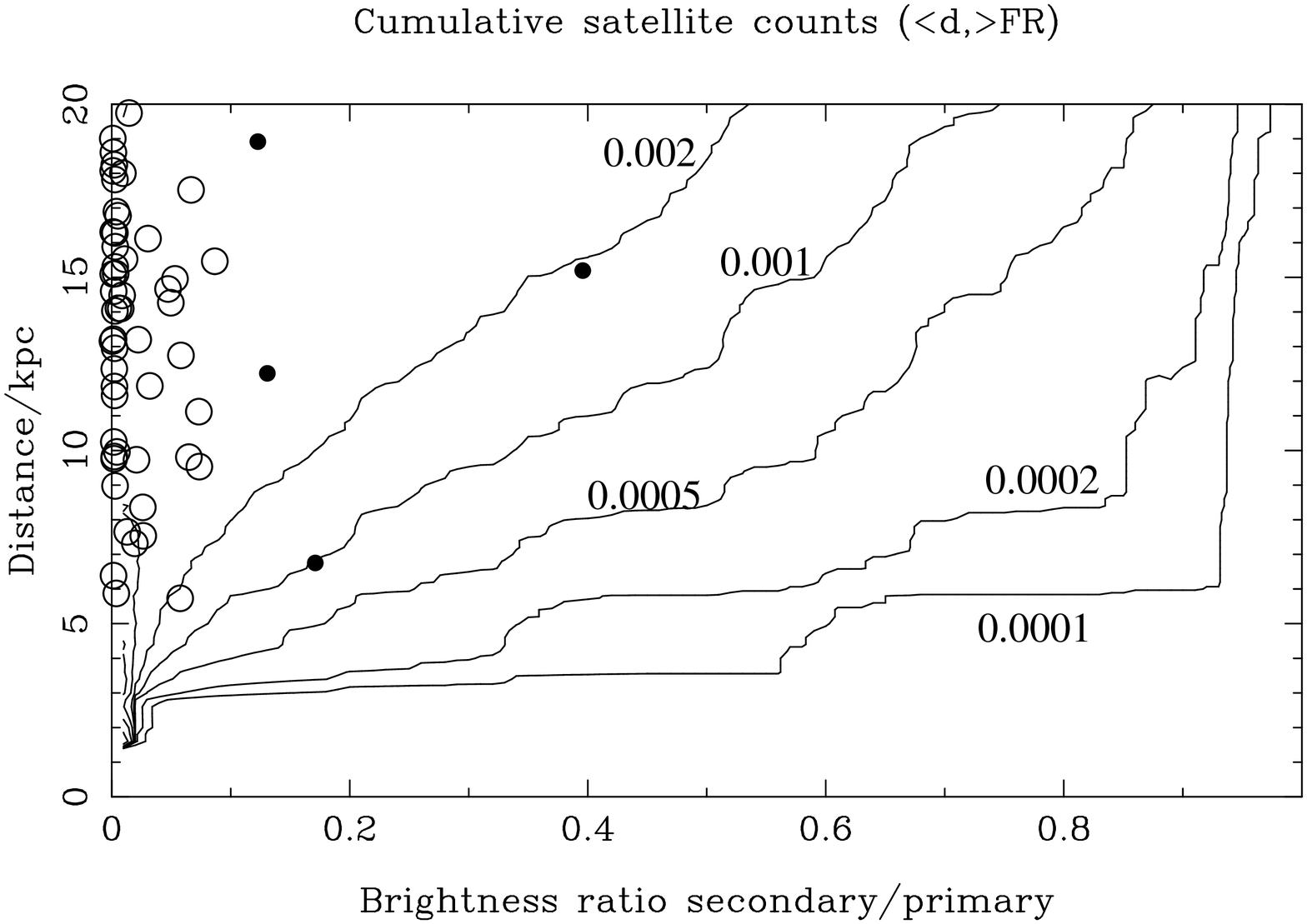}
\caption{Left: secondaries detected in the SLACS images within 7\arcsec\,
of each primary, plotted against $I$ magnitude and distance from the
primary. Contours indicate the likelihood at each point of the satellite
being a random background object, based on the COSMOS source counts. The
four satellites for which the secondary flux is more than 0.1 times that
of the primary are indicated by solid circles. Right: SLACS secondaries,
plotted on a graph of secondary/primary flux ratio against linear
distance. The contours represent the background-subtracted COSMOS counts
as in Fig. 4. Note the lack of obvious close, high flux-ratio
secondaries as seen in the CLASS survey. }
\label{figslacs}
\end{figure*}

Over most of the range of brightness and angular distance, the
distribution of SLACS satellites is statistically indistinguishable from
background (for example, 29 satellites are expected with $I_{814}<24$,
$d<6\arcsec$ and 34 are observed, with a further 3-4 probably missed due
to limited sensitivity). A few satellites are likely to be genuinely
associated with the primary, including for example the satellite of
SLACS~0808+4706; this object lies at a cumulative probability of 0.005,
implying that only 0.3 satellites of this brightness and distance should
occur in the SLACS sample. Nevertheless, these satellites do not lie in
the region of flux ratio-linear distance space that is strongly
discrepant with the COSMOS statistics (right hand side of Fig. 11) as
we see in the CLASS images.

\section{Discussion and conclusions}

The current work sharpens the problem which arises from the presence of
excess satellites around lensing galaxies in the CLASS survey. It
appears that the frequency of satellite galaxies, predicted by current 
simulations based on CDM/semi-analytic models are in excess of
the actual frequency of satellites observed in the COSMOS survey and
also in the wider but shallower SDSS survey,
given the limits imposed by resolution and depth. Though the sample
statistics are still small, lensing galaxies in the SLACS survey are
also consistent with early-type, non-lensing galaxies in COSMOS. This is
interesting as the SLACS survey lens-galaxy sample has a much lower
median redshift (z=0.20) than the COSMOS survey galaxies. This leaves
the CLASS lens galaxies as the anomalies, containing significantly
higher rates of luminous substructure than either simulations or field
galaxies in SDSS and COSMOS. The major significant
difference between CLASS and SLACS lenses is that the redshift of lens
(and source) is typically larger in the CLASS lenses. However, the
difficulty then becomes the difference between lensing galaxies (CLASS)
and non-lensing galaxies (COSMOS) in broadly similar redshift ranges.

One possibility is that, although the incidence of satellites in the
CLASS lens systems appears anomalously high, the effect may still be due
to small-sample statistics. An obvious priority is the acquisition of
larger samples of high-redshift lenses. Although existing radio surveys
can be done more efficiently (Jackson \& Browne 2007) major progress
probably awaits future instruments such as LOFAR and the SKA (Koopmans,
Browne \& Jackson 2004). Optical surveys capable of detecting
large numbers of high-redshift galaxy lens systems are a few years in the
future using telescopes such as Pan-STARRS and particularly the LSST.

In semi-analytic models, the satellites which are found within
$\sim$10~kpc of the central galaxy correspond to sub-haloes which are
tidally stripped of their dark halo during repeated orbits around the primary
galaxy. It is natural to ascribe the close satellites in CLASS lens
systems to these ``orphan'' galaxies, but there are still problems with
this approach and recent simulations using Millenium-II may reproduce
the observed clustering without orphans (Guo et al. 2009). There would 
need to be a redshift dependence (Bryan et al. 2008) which allowed us 
to see them in the high-redshift CLASS lenses but not the
low-redshift SLACS lenses, and they would need to increase the lensing
cross section sufficiently to bias the lensing statistics. It is hard to
imagine how this happens; many studies (e.g. Chen et al. 2003, Metcalf
\& Zhao 2002) have addressed the question of how satellites could produce flux
ratio anomalies, but it seems unlikely that
the extra magnifications produced would account for the anomalous
statistics in the CLASS lens systems. Alternatively, the fact that the
typical CLASS galaxy is brighter than the typical COSMOS object by just
over a magnitude (Fig. 1) could be a reflection of CLASS galaxies being
hosted by more massive dark matter haloes. The simulations (Section 2.4)
imply that a difference of a factor 10 in dark matter halo is needed to
change the incidence of close satellites by a factor of 2-3, so such an
effect, while present, may not be enough to account for the full
difference between CLASS and COSMOS.

The comparison with simulations would benefit a great deal from the use of
higher resolution simulations involving a realistic treatment of the gas
processes, but we highlight the fact that studying the central regions of
haloes may provide important additional constraints to theoretical models. 
This is because the models may reproduce the overall luminosity function of
(more numerous) satellite galaxies, but fail for the small fraction of 
projected central galaxies which are most sensitive to numerical effects
(e.g. resolutions) and physical processes such as tidal stripping.

\section*{Acknowledgements}

This research was supported by the EU Framework 6 Marie Curie Early
Stage Training programme under contract number MEST-CT-2005-19669
``ESTRELA''. We thank the SLACS team, especially Leon Koopmans and Adam
Bolton, for the provision of images of SLACS lenses, together with mask
images, which were used in this work; and Leon Koopmans for useful
discussions. We thank Gerard Lemson for providing simulated galaxy
catalogues used in this work. In addition, we thank the referee for 
useful comments. 
The Millennium Simulation databases used in this paper and
the web application providing online access to them were
constructed as part of the activities of the German Astrophysical
Virtual Observatory.
Funding  for the  SDSS and  SDSS-II has  been provided  by  the Alfred
P.  Sloan  Foundation, the  Participating  Institutions, the  National
Science  Foundation,  the  U.S.  Department of  Energy,  the  National
Aeronautics and Space Administration, the Japanese Monbukagakusho, the
Max  Planck Society,  and  the Higher  Education  Funding Council  for
England. The SDSS Web Site is http://www.sdss.org/. The SDSS is  managed 
by the Astrophysical Research  Consortium for the
Participating  Institutions. The  Participating  Institutions are  the
American Museum  of Natural History,  Astrophysical Institute Potsdam,
University  of Basel,  University of  Cambridge, Case  Western Reserve
University,  University of Chicago,  Drexel University,  Fermilab, the
Institute  for Advanced  Study, the  Japan Participation  Group, Johns
Hopkins University, the Joint  Institute for Nuclear Astrophysics, the
Kavli Institute  for Particle  Astrophysics and Cosmology,  the Korean
Scientist Group, the Chinese  Academy of Sciences (LAMOST), Los Alamos
National  Laboratory, the  Max-Planck-Institute for  Astronomy (MPIA),
the  Max-Planck-Institute  for Astrophysics  (MPA),  New Mexico  State
University,   Ohio  State   University,   University  of   Pittsburgh,
University  of  Portsmouth, Princeton  University,  the United  States
Naval Observatory, and the University of Washington.

\baselineskip 10pt

\section*{References}
\parskip=0.15cm

\noindent Adelman-McCarthy J.K., et al., 2008, ApJS 175, 297

\noindent Augusto P., et al., 2001, MNRAS 326, 1007

\noindent Belokurov V., Zucker D.B., Evans N.W., Wilkinson M.I., Irwin M.J., Hodgkin S., Bramich D.M., Irwin J.M., Gilmore G., Willman B.,et al. 2006,  ApJ, 647, L111. 

\noindent Belokurov V., Zucker D.B., Evans N.W., Kleyna J.T., Koposov S., Hodgkin S.T., Irwin M.J., Gilmore G., Wilkinson M.I., Fellhauer M.,et al. 2007,  ApJ, 654, 897. 

\noindent Bertin E., Arnouts S. 1996,  A\&AS, 117, 393. 

\noindent Biggs A.D., Browne I.W.A., Jackson N.J., York T., Norbury M.A., McKean J.P., Phillips P.M. 2004,  MNRAS, 350, 949. 

\noindent Blanton M.R., et al., 2005, AJ 129, 2562

\noindent Blanton M.R., Roweis S., 2007, AJ 133, 734

\noindent Bolton A.S., et al., 2006, ApJ 638, 703

\noindent Bolton A.S., et al., 2008, ApJ 682, 964

\noindent Browne I.W.A., et al., 2003, MNRAS 341, 13

\noindent Bryan S., Mao S., Kay S.T., 2008, MNRAS 391, 959

\noindent Bullock J.S., Kravtsov A.V., Weinberg D.H. 2000,  ApJ, 539, 517. 

\noindent Capak P., Aussel H., Ajiki M., McCracken H.J., Mobasher B., Scoville N., Shopbell P., Taniguchi Y., Thompson D., Tribiano S.,et al. 2007,  ApJS, 172, 99. 

\noindent Chen J. 2009, A\&A 494, 867

\noindent Chen J., Kravtsov A.V., Keeton C.R., 2003, ApJ 592, 24

\noindent Chen J., Kravtsov A.V., Prada F., Sheldon E., Klypin A.A.,
Blanton M.R., Brinkmann J., Thakar A.R., 2006, ApJ 647, 86

\noindent Chiba M. 2002,  ApJ, 565, 17. 

\noindent Cohn J., Kochanek C.S., 2004, ApJ, 608, 25. 

\noindent Cole S., et al. 2005,  MNRAS, 362, 505. 

\noindent Dalal N., Kochanek C.S. 2002,  ApJ, 572, 25. 

\noindent De Lucia G., Blaizot J. 2007, MNRAS, 375, 2.

\noindent Diemand J., Kuhlen M., Madau P. 2007,  ApJ, 667, 859. 

\noindent Efstathiou G., 1992, MNRAS, 256, P43

\noindent Eisenstein D.J., et al. 2005,  ApJ, 633, 560. 

\noindent Falco E.E., Leh\'ar J., Shapiro I.I., 1997, AJ 113, 540

\noindent Faure C., et al., 2008, ApJS 176, 19

\noindent Gao L., White S.D.M., Jenkins A., Stoehr F., Springel V. 2004,  MNRAS, 355, 819. 

\noindent Ghigna S., Moore B., Governato F., Lake G., Quinn T., Stadel
J., 2000, ApJ, 544, 616

\noindent Gnedin N. Y., 2000, ApJ, 542, 535	

\noindent Guo Q., White S., Li C., Boylan-Kolchin M., 2009, astro-ph/0909.4305

\noindent Impey, C.D., Foltz, C.B., Petry, C.E., Browne, I.W.A., Patnaik,
A.R., 1996, ApJ 462, L53

\noindent Jackson N., Nair S., Browne I.W.A., 1998, in ``Observational Cosmology
with the new radio surveys'', Astrophysics \& Space Science Library vol 226,
eds. Jackson N. et al., publ. Kluwer, Dordrecht

\noindent Jackson N., Browne I.W.A., 2007, MNRAS 374, 168

\noindent Jackson N., Xanthopoulos E., Browne I.W.A., 2000, MNRAS 311, 389

\noindent Jackson N., 2008, MNRAS 389, 1311

\noindent Jaunsen A.O., Hjorth J., 1997, A\&A 317, L39

\noindent King L.J., et al., 1998, MNRAS 295, L41

\noindent Klypin A., Kravtsov A.V., Valenzuela O., Prada F. 1999,  ApJ, 522, 82. 

\noindent Kochanek C.~S., 1991, ApJ, 373, 354

\noindent Kochanek C.S., Dalal N. 2004,  ApJ, 610, 69. 

\noindent Koekemoer A.M., Aussel H., Calzetti D., Capak P., Giavalisco M., Kneib J.P., Leauthaud A., LeF\'evre O., McCracken H.J., Massey R.,et al. 2007,  ApJS, 172, 196. 

\noindent Koopmans L.V.E., et al., 1999, MNRAS 303, 727

\noindent Koopmans L.V.E., Browne I.W.A., Jackson N., 2004, NewAR 48, 1085

\noindent Lawrence C.R., Neugebauer G., Matthews K. 1993,  AJ, 105, 17. 

\noindent Li C., Kauffmann G., Wang L., White S.D.M., Heckman T.M., Jing
Y.P., 2006, MNRAS 373, 457

\noindent Li C., Kauffmann G., Heckman T.M., Jing Y.P., White S.D.M.,
2008a, MNRAS 385, 1903

\noindent Li C., et al., 2008b, MNRAS 385, 1915

\noindent Madau P., Diemand J., Kuhlen M. 2008,  ApJ, 679, 1260. 

\noindent Mao S., Schneider P. 1998,  MNRAS, 295, 587. 

\noindent {Mao} S.,  {Jing} Y.,  {Ostriker} J.~P.,    {Weller} J.,  2004, ApJ, 604, L5
	
\noindent McKean J.P., Koopmans, L.V.E., Browne, I.W.A., Fassnacht,
C.D., Blandford R.D., Lubin, L.M., Readhead, A.C.S., 2004, MNRAS 350, 167

\noindent McKean J.P., Koopmans L.V.E., Flack C.E., Fassnacht C.D., Thompson D., Matthews K., Blandford R.D., Readhead A.C.S., Soifer B.T. 2007,  MNRAS, 378, 109. 

\noindent Metcalf R.B. 2002,  ApJ, 580, 696. 

\noindent Metcalf R.B., Zhao H., 2002, ApJ 567, L5

\noindent Moore B., Ghigna S., Governato F., Lake G., Quinn T., Stadel J., Tozzi P. 1999,  ApJ, 524, L19. 

\noindent More A., McKean J.P., More S., Porcas R.W., Koopmans L.V.E., Garrett M.A., 2009, MNRAS 394, 174

\noindent Myers S.T., et al., 2003, MNRAS 341, 1

\noindent Percival W., et al., 2001, MNRAS 327, 1297.

\noindent Rusin D., Norbury, M., Biggs A.D., Marlow, D.R., Jackson, N.
J., Browne I.W.A., Wilkinson P.N., Myers S.T., 2002, MNRAS 330, 205

\noindent Rusin E., et al., 2001, ApJ 557, 594

\noindent Schechter P.L., Moore C.B. 1993,  AJ, 105, 1. 

\noindent Schneider D.P., Lawrence C.R., Schmidt M., Gunn J.E., Turner E.L., Burke B.F., Dhawan V. 1985,  ApJ, 294, 66. 

\noindent Scoville N., Abraham R.G., Aussel H., Barnes J.E., Benson A., Blain A.W., Calzetti D., Comastri A., Capak P., Carilli C.,et al. 2007,  ApJS, 172, 38. 

\noindent Shin E.M., Evans N.W. 2008,  MNRAS, 385, 2107. 

\noindent Spergel D.N., et al. 2007,  ApJS, 170, 377. 

\noindent Springel V., et al., Nature 435, 629

\noindent {Springel} V.,  et al., 2008,  MNRAS 391, 1685

\noindent Suyu, S.H., Marshall, P.J., Blandford, R. D., Fassnacht, C.D., Koopmans, L.V.E., McKean, J.P., Treu,
T., 2009, ApJ 691, 277

\noindent Tegmark M., et al. 2004,  ApJ, 606, 702. 

\noindent {Thoul}, A.~A. and {Weinberg}, D.~H., 1996, ApJ, 465, 608

\noindent Willman B., Dalcanton J.J., Martinez-Delgado D., West A.A., Blanton M.R., Hogg D.W., Barentine J.C., Brewington H.J., Harvanek M., Kleinman S.J.,et al. 2005,  ApJ, 626, L851. 

\noindent Xu, D. D., Mao, S., Wang, J., Springel, V., Gao, L., White,
S. D. M., Frenk, C. S., Jenkins, A., Li, G. L., Navarro, J. F. 2009,
MNRAS, 398, 1235

\noindent York D.G., Adelman J., Anderson J.E. J., Anderson S.F., Annis J., Bahcall N.A., Bakken J.A., Barkhouser R., Bastian S., Berman E.,et al. 2000,  AJ, 120, 1579. 

\noindent York T., Jackson N., Browne I.W.A., Wucknitz O., Skelton J.E., 2005,
MNRAS 357, 124

\noindent York T., et al., 2005, MNRAS 361, 259

\noindent Zucker D.B., Belokurov V., Evans N.W., Kleyna J.T., Irwin M.J., Wilkinson M.I., Fellhauer M., Bramich D.M., Gilmore G., Newberg H.J.,et al. 2006,  ApJ, 650, L41. 

\noindent Zucker D.B., Belokurov V., Evans N.W., Wilkinson M.I., Irwin M.J., Sivarani T., Hodgkin S., Bramich D.M., Irwin J.M., Gilmore G.,et al. 2006,  ApJ, 643, L103. 

\end{document}